\documentclass[submission,copyright,creativecommons]{eptcs}
\providecommand{\event}{WRS 2011} 
\usepackage{graphicx}
\usepackage{latexsym}
\usepackage{alltt}

\newcommand{\codesize}{\fontsize{10pt}{12pt}\selectfont}

\newcommand{\at}{\mathbin{@}}
\newcommand{\bigfrac}[2]{\displaystyle\frac{#1}{#2}}

\title{Basic completion strategies as another application of the Maude strategy language}
\author{Alberto Verdejo \qquad\qquad Narciso Mart\'\i-Oliet
\institute{Facultad de Inform\'atica\\
Universidad Complutense de Madrid\\
Madrid, Spain}
\email{\quad alberto@sip.ucm.es \qquad\qquad narciso@esi.ucm.es}
}
\def\titlerunning{Basic completion strategies as another application of the Maude strategy language}
\def\authorrunning{A. Verdejo \& N. Mart\'\i-Oliet}

\begin{document}
\maketitle

\begin{abstract}
The two levels of data and actions on those data provided by the separation between equations
and rules in rewriting logic are completed by a third level of strategies to control the application
of those actions.  This level is implemented on top of Maude as a strategy language, which has
been successfully used in a wide range of applications. First we summarize the 
Maude strategy language design and review some of its applications; then, we describe a 
new case study, namely the description of \emph{completion procedures 
as transition rules + control}, as proposed by Lescanne.
\end{abstract}

\section{Introduction}

Strategies are pervasive in Computer Science; we have, among many others, control strategies,
reduction strategies, deduction strategies, rewriting strategies, narrowing strategies, theorem-proving
strategies, e-learning strategies, etc. This is just a reflection of the fact that strategies are essential
ingredients in game and problem-solving.  In general, in many settings we can identify three levels 
in the development of the solution for a given problem:
\begin{itemize}
\item  definition of the data involved in the problem,
\item  identification of the basic actions that manipulate those data, and
\item  strategies to specify how those basic actions must be used to reach the desired solution. 
\end{itemize}
For example, in the setting of business process modeling, those three levels correspond to the 
data (clients, reservations, money, etc.), the business activities or web services involved, and 
the composition of such activities or services to design more complex interactions through 
languages such as BPEL \cite{BPEL} and BPMN \cite{BPMN}.

Since its introduction by Meseguer  in the early nineties \cite{tcs/Meseguer92}, rewriting logic 
addressed the separation between the first two levels above by distinguishing at the logic level
equations from rules.  Equations are used to define data (possibly including states), while rules 
are used to define transitions, activities, actions, and so on, that use data and allow to move from 
one state to another.  Different specification languages are directly based on rewriting logic,
including ELAN \cite{BKKR01,BorovanskyKirchnerKirchnerMoreau02} and Maude 
\cite{maude/2007,CDELMOMT11}; in those languages, the distinction between equations and
rules is emphasized by requiring, although both are implemented in terms of rewriting,
equations to be confluent and terminating (and thus, any reduction strategy 
will give rise to the same unique result), while rules need not be either confluent or terminating.
Then, if the user is interested in controlling the application of rules to avoid undesired 
directions, either the control is introduced into the rules in an ad hoc way depending on
the problem, or it is necessary to introduce somehow the third level above to
control the rule application by means of strategies.  In the case of ELAN, this was part of its design, so that 
strategies become an essential part of the ELAN system, which
provides a basic set of strategies which can be used when writing rewrite
rules, in such a way that at the specification level it is not enforced a separation
between rules and strategies.   On the other hand, in the case of Maude, for a while the
introduction of explicit strategies was avoided by means of its direct access to the metalevel.
Indeed, the Maude system provides \texttt{rewrite} commands for getting only an execution path,
as well as a \texttt{search} command for exploring all possible execution paths from 
a starting term \cite{CDELMOMT11}; if one is interested in the results of only some 
execution paths satisfying some constraints, these can typically be specified at the
metalevel, where both equations and rules become simply more data and can be
manipulated in different ways. Even more, several strategy languages at the metalevel
have been considered for different applications, such as, for example \cite{Clavel-Meseguer97}
for completion. 

Taking into account our own previous experience
designing strategy languages in Maude, and also from the experience of
other languages like the already mentioned ELAN, 
TOM \cite{TOM-manual}, and Stratego \cite{Visser01,Visser04} 
we decided to design a strategy language for Maude \cite{MartiOlietMeseguerVerdejo04}, 
to be used \emph{at the object level} instead of at the metalevel, thus avoiding the need
to know this more complex framework, and completing at the same time in the case of
Maude the third level in problem solving as described above.

The Maude strategy language allows the definition of strategy expressions that
control the way a term is rewritten.  Differently from ELAN, our design was based 
on a strict separation between the
rewrite rules and the strategy expressions, that are provided in separate
modules. Thus, in our language it is not possible to use strategy
expressions in the rewrite rules of a system module.
A strategy is described as an operation that, when applied to a given term,
produces a set of terms as a result, given that the process is
nondeterministic in general. The basic strategies consist of the application of 
a rule (identified by the corresponding rule label) to a given term, and allowing 
variables in a rule to be
instantiated before its application by means of a substitution. For conditional rules, 
rewrite conditions can be controlled also by means of
strategies. Basic strategies are combined by means of, among others, 
typical regular expression 
constructions (concatenation, union, and iteration), if-then-else, combinators to
control the way subterms of a given term are rewritten, and recursion 
\cite{MartiOlietMeseguerVerdejo04}. 

Since its proposal, the language has been successfully applied to a wide range of 
examples, from operational semantics representations to the formalization of web services;
in particular, in the context of business process modelling that we have mentioned
above, the Maude strategy language has been used to represent in Maude fragments
of BPEL \cite{iswsa/MerouaniMS10} and also of BPMN \cite{Henche10}.

In the first part of this paper, after a quick introduction to Maude, we summarize the 
Maude strategy language design, by reviewing its 
syntax and set-theoretic semantics in Section~\ref{sec:strat-lang}, and then survey 
the main applications in Section~\ref{sec:strat-apps}. In the second part of the paper,
Section~\ref{sec:completion-strats}, we present a new case study, namely the 
description of \emph{completion procedures 
as transition rules + control}, as proposed by Lescanne in \cite{Lescanne89}. 
Equational systems are represented as data
that is going to be manipulated by rewrite rules implementing the completion
inference rules, following a well-known approach. In order to get a completion
algorithm, one needs to apply these rules in a controlled way. In his paper,
Lescanne does this by means of CAML programs, while our approach is more
abstract, emphasizing the fact that inference rules do not change at all in the
different algorithms. 

\section{Maude in a nutshell}\label{sec:maude}

In Maude the state of a system is formally specified as an
algebraic data type by means of an equational specification. Maude
uses a very expressive version of equational logic, namely
\emph{membership equational logic}
\cite{tcs/BouhoulaJM00}. In this kind of specifications
we can define new types (by means of the keyword
\texttt{sort}); subtype relations between types
(\texttt{subsort}); operators (\texttt{op}) for building values of
these types, giving the types of their arguments and result, and
which may have attributes as being associative (\texttt{assoc}) or
commutative (\texttt{comm}), for example; equations (\texttt{eq})
that identify terms built with these operators; and memberships
(\texttt{mb}) $t:s$ stating that the term $t$ has sort $s$. Both
equations and memberships can be conditional. Conditions are
formed by a conjunction (written \verb+/\+) of equations and
memberships.
%
%
%
Equations are assumed to be confluent and terminating, that is, we
can use the equations from left to right to reduce a term $t$ to a
unique (modulo the operator attributes such as associativity or
commutativity) canonical form $t'$  that is
equivalent to $t$, i.e.\ they represent the same value.

The \emph{dynamic} behavior of a system is specified by rewrite
rules of the form
\[\vspace*{-1ex}
l:\ t \longrightarrow t' \ {\it if} \;\; (\bigwedge_{i}
u_{i}=v_{i}) \wedge(\bigwedge_{j} w_{j}:s_{j}) \wedge
(\bigwedge_{k} p_{k}\longrightarrow q_{k})
\vspace*{-1ex}\]
that describe the local, concurrent transitions of the
system. That is, when part of a system matches the pattern $t$ and
the conditions are fulfilled, it can be transformed into the
corresponding instance of the pattern~$t'$.

\subsection{Crossing the river}
\label{ex:river-cross}

This example is taken from \cite[Section~7.8]{maude/2007}, although the presentation
here is slightly different.

A shepherd needs to transport to the other side of a river a wolf, a goat, and a
cabbage. He has only a boat with room for the shepherd himself and another item.
The problem is that in the absence of the shepherd the wolf would eat the goat, and
the goat would eat the cabbage.

We represent with constants \texttt{left} and \texttt{right} the two sides of the
river. The shepherd and his belongings are represented as objects with an attribute
indicating the side of the river in which each is located and are grouped together 
with a multiset operator \verb|__|. The rules represent how the wolf or the goat eat and 
the ways of crossing the river allowed by the 
capacity of the boat; an auxiliary \texttt{change} operation is used
to modify the corresponding attributes.

{\codesize
\begin{verbatim}
mod RIVER-CROSSING is
  sorts Side Group .
  ops left right : -> Side .
  op change : Side -> Side .
  ops s w g c : Side -> Group .  --- shepherd, wolf, goat, cabbage
  op __ : Group Group -> Group [assoc comm] .

  vars S S' : Side .
  eq change(left) = right .
  eq change(right) = left .

  crl [wolf-eats] : w(S) g(S) s(S') => w(S) s(S') if S =/= S' .
  crl [goat-eats] : c(S) g(S) s(S') => g(S) s(S') if S =/= S' .
  rl [shepherd-alone] : s(S) => s(change(S)) .
  rl [wolf] : s(S) w(S) => s(change(S)) w(change(S)) .
  rl [goat] : s(S) g(S) => s(change(S)) g(change(S)) .
  rl [cabbage] : s(S) c(S) => s(change(S)) c(change(S)) .
endm
\end{verbatim}
}

We want to know if there is a way
the shepherd can safely take his belongings to the other side. 
But if we search if a state where everybody is on the right is
reachable from a state where everybody is on the left, we cannot
be sure that an intermediate state where, for example, the wolf
has the posibility of eating but it has not eaten, is also reached.
That is, with the rewrite or search commands of Maude we cannot
ensure that the priority of the first two rules is taken into account.
In Section~\ref{ex:river-cross-strat} we will present a strategy 
that ensures that priority.

\section{The Maude strategy language}
\label{sec:strat-lang}

In this section we describe the syntax of our strategy
language and its set-theoretic semantics. A strategy is described as an operation that, when applied to a given term,
produces a \emph{set} of terms as a result, given that the process is
nondeterministic in general. If the strategy \emph{fails} (it cannot be applied to the given term),
the empty set of results is returned. Otherwise, we say that the strategy \emph{succeeds} 
(possibly returning several results). For the strategies of a  rewrite theory
$\mathcal{R}$ with signature $\Sigma$ we have a function
\[ \_@\_ : Strat \times T_{\Sigma}(X) \longrightarrow  
  \mathcal{P}( T_{\Sigma}(X)),
 \]
where $T_{\Sigma}(X)$ denotes the set of $\Sigma$-terms with variables in $X$.
This function has an obvious extension to a function
\[ \_@\_ : Strat \times \mathcal{P}(T_{\Sigma}(X)) \longrightarrow  
  \mathcal{P}( T_{\Sigma}(X)),
 \]
where, if $\sigma \in Strat$ and $U \subseteq T_{\Sigma}(X)$,
 we have $\sigma \at U = \bigcup_{t \in U} \sigma \at t$.

\subsection{Idle and fail}

The simplest strategies are the constants \texttt{idle} and
\texttt{fail}.  The first always succeeds, but without modifying the
term $t$ to which it is applied, that is, $\texttt{idle}\at t=\{t\}$,
while the second always fails, that is, $\texttt{fail}\at t=\emptyset$.

\subsection{Basic strategies}

The basic strategies consist of the application of a rule (identified
by the corresponding rule label) to a given term. 
Rule variables can be
instantiated before its application by means of a substitution, that is, a 
mapping of variables to terms, so that the
user has more control on the way the rule is applied.
In case of conditional rules, the default
breadth-first search strategy is used for checking the rewrites in the
condition.  Therefore, if $l$ is a rule label, $s$ a substitution, and $t$ a term,
the semantics of $l\texttt{[}s\texttt{]}\at t$ is the set of terms to which $t$ rewrites in
one step using the rule with label $l$ instantiated by substitution $s$
\emph{anywhere} where it matches
and satisfies the rule's condition. The substitution can be omitted if it is empty.

For conditional rules, rewrite conditions can be controlled by means of
strategy expressions. A strategy expression of the form $l\verb"["s\verb"]{" \sigma_1 \ldots \sigma_n\verb"}"$ 
denotes a basic strategy that
applies \emph{anywhere} in a given state term the rule $l$ with variables
instantiated by means of the substitution $s$ and using $\sigma_1, \ldots, \sigma_n$ as strategy expressions to check the rewrites in the
condition of $l$. The number of rewrite
condition fragments appearing in the condition of rule 
$l$ must be exactly $n$ for the
expression to be meaningful. 

\subsection{Top}

The most common case allows applying a rule \emph{anywhere} in a
given term, as explained above, but we also provide an operation to
restrict the application of a rule only to the \emph{top} of the term,
because in some examples like structural operational semantics, the
only interesting or allowed rewrite steps happen at the top.
\verb"top("$\beta$\verb")" applies the basic strategy $\beta$ only at the top 
of a given state term. Note, however, that even applying a rule at the top is
nondeterministic due to the possibility of multiple matches,  because
matching takes place modulo the equational attributes of the operators, such
as associativity, commutativity, or identity.

\subsection{Tests}

Tests are considered as strategies that check a property on a state, so that
the strategy applied to a state is successful when the test is true on such a state,
and the strategy fails when the test is false; moreover,
in the first case the state is not changed.  That is, for $\tau$ a test
and $t$ a term, $\tau\at t$ will evaluate to $\{t\}$ if $\tau$ succeeds on $t$,
and to $\emptyset$ if it fails, so that $\tau$ acts as a filter on its
input.

Since matching is one of the basic steps that take place when applying a
rule, the strategies that test some property of a given state term are based on
matching. As in applying a rule, we distinguish between matching anywhere
and matching only at the top of a given term. 
\verb"amatch "$\rho$\verb" s.t." $C$ is a test that, when applied to a given state term
$t'$, is successful if there is a subterm of $t'$ that matches the
pattern $\rho$ (that is, matching is allowed \emph{anywhere} in the state
term) and then the condition $C$ is satisfied with the substitution
for the variables obtained in the matching, and fails otherwise.
\verb"match "$\rho$\verb" s.t." $C$ corresponds to matching only at the \emph{top}.
When the condition $C$ is simply \texttt{true}, it can be omitted.

\subsection{Regular expressions} \label{regular}

Basic strategies can be combined so that strategies are applied to execution paths.
The first strategy combinators we consider are the typical regular expression 
constructions: concatenation, union, and iteration. 
The concatenation operator is associative and the union operator is
associative and conmutative.  This commutativity of union provides a
form of nondeterminism in the way the solutions are found.

If $\sigma$, $\sigma'$ are strategy expressions and $t$ is a term,
then $(\sigma \,\texttt{;}\, \sigma')\at t=\sigma'\at  (\sigma\at t)$,
$(\sigma \,\texttt{|}\, \sigma')\at t=(\sigma\at t) \cup (\sigma'\at t)$, and 
$\sigma~\texttt{+}\at t= \bigcup_{i \geq 1} \sigma^i \at t$, where
$\sigma^1 = \sigma$ and $\sigma^n = (\sigma \,\texttt{;}\, \sigma^{n-1})$ for $n > 1$.
Of course, $\sigma~\texttt{*}=\texttt{idle}\,\texttt{|}\, \sigma~\texttt{+}$.
For example, a strategy of the form $\sigma \,\texttt{;}\, \tau$ (with $\tau$ a
test) will filter out
all those results from $\sigma$ that do not satisfy the test $\tau$.

\subsection{Conditional strategy and its derivations} \label{if-then-else}

Our next strategy combinator is a typical if-then-else, but generalized so
that the first argument is also a strategy following ideas from
Stratego \cite{Visser04} and ELAN \cite{ijfcs/BorovanskyKKR01}.

The behavior of the strategy expression $\sigma \;\texttt{?}\; \sigma' \;\texttt{:}\; \sigma''$ is
as follows: in a given state term, the strategy $\sigma$ is evaluated; if
$\sigma$ is successful, the strategy $\sigma'$ is evaluated in the
resulting states, otherwise $\sigma''$ is evaluated in the \emph{initial}
state.  That is
\[(\sigma \;\texttt{?}\; \sigma' \;\texttt{:}\; \sigma'')\at t = \mathbf{if} \; (\sigma\at t) \not= \emptyset \; \mathbf{then} \;
\sigma'\at (\sigma\at t) \; \mathbf{else} \; \sigma''\at t \; \mathbf{fi}.\] Note that, as
mentioned above, in general the first argument is a strategy
expression and not just a test.  Since a test is a strategy,
we have the particular case $\tau \;\texttt{?}\; \sigma' \;\texttt{:}\; \sigma''$ for a
test $\tau$ where evaluation coincides with the typical Boolean
case distinction: $\sigma'$ is evaluated when the test $\tau$ is
true and $\sigma''$ when the test is false, taking into account that
a test fails when false.

Using the conditional combinator, we can define many other useful strategy
combinators as derived operations. $\sigma \;\texttt{orelse}\; \sigma'$ evaluates
$\sigma$ in a given state; if such evaluation is successful, its results
are the final ones, but if it fails, then $\sigma'$ is evaluated in the
initial state.
\[ \sigma \;\texttt{orelse}\; \sigma' = 
   \sigma \;\texttt{?}\; \texttt{idle} \;\texttt{:}\; \sigma' \]

\texttt{not($\sigma$)} reverses the result of evaluating $\sigma$, so that
\texttt{not($\sigma$)} fails when $\sigma$ is successful and vice versa.
\[ \texttt{not($\sigma$)} = \sigma \;\texttt{?}\; \texttt{fail} \;\texttt{:}\; \texttt{idle}\]
An interesting use of \texttt{not($\sigma$)} is the following ``normalization''
(or ``repeat until the end'') operation $\sigma$~\texttt{!}:
\[ \sigma~\texttt{!} = \sigma~\texttt{* ; not(}\sigma\texttt{)}\]

\texttt{try($\sigma$)} evaluates $\sigma$ in a given state; if it is successful,
the corresponding result is given, but if it fails, the initial state is 
returned.
\[ \texttt{try($\sigma$)} = \sigma \;\texttt{?}\; \texttt{idle} \;\texttt{:}\; \texttt{idle}\]

Evaluation of \texttt{test($\sigma$)} checks the success/failure result of
$\sigma$, but it does not change the given initial state.
\[ \texttt{test($\sigma$)} = \texttt{not($\sigma$)} \;\texttt{?}\; \texttt{fail} \;\texttt{:}\; \texttt{idle}\]
Notice that $\texttt{test($\sigma$)} = \texttt{not(not($\sigma$))}$.

\subsection{Rewriting of subterms}
\label{xmatchrew}

With the previous combinators we cannot force the application of a
strategy to a specific subterm of the given initial term. 
We can have more control over the way
different subterms of a given state are rewritten by means of the
\texttt{amatchrew} combinator.

\begin{figure}
\begin{center}
\includegraphics{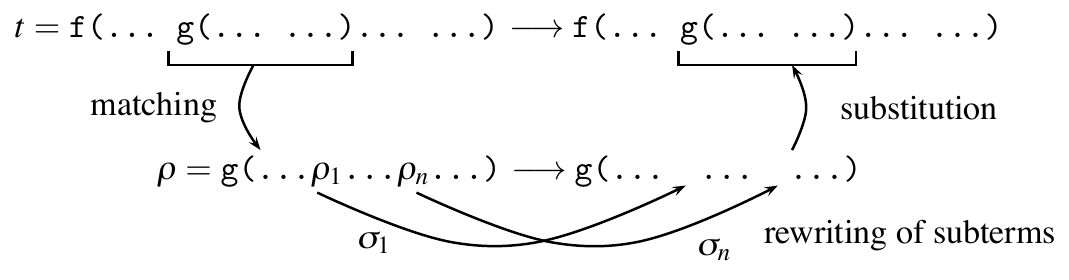}
\end{center}
\vspace*{-1em}
\caption{Behavior of the \texttt{amatchrew} combinator.}\label{xmatchrew-figure}
\end{figure}

When the strategy 
\begin{center}
\texttt{amatchrew $\rho$ s.t.\ $C$ by $\rho_1$ using 
$\sigma_1$,$\ldots$, $\rho_n$ using $\sigma_n$}
\end{center}
is applied to a state term $t$, first a subterm of
$t$ that matches the pattern $\rho$ and satisfies $C$ is selected. Then,
the patterns $\rho_1, \ldots, \rho_n$ (which must be disjoint subterms of
$\rho$), instantiated appropriately, are rewritten as described by the
strategy expressions $\sigma_1, \ldots, \sigma_n$, respectively. The results 
are combined in $\rho$ and then substituted in $t$, 
in the way illustrated in Figure~\ref{xmatchrew-figure}.

The strategy expressions $\sigma_1, \ldots, \sigma_n$ can make use of the
variables instantiated in the matching, thus taking care of information
extracted from the state term (see, for example, the strategies in
Section~\ref{Ncompletion}).

The version \texttt{matchrew} works in the same way, but performing
matching only at the top. In both cases, the condition can be omitted
when it is \texttt{true}.

The \emph{congruence operators} used in ELAN and Stratego \cite{ijfcs/BorovanskyKKR01,Visser04}
are special cases of the \texttt{matchrew} combinator, as shown in 
\cite{MartiOlietMeseguerVerdejo04}.

\subsection{Strategy modules and commands}
\label{strat-modules}

Given a Maude system module $M$, the user can write one or more
strategy modules to define strategies for $M$. 
Such strategy modules have the following form:

{\codesize
\begin{alltt}
smod \(\mathit{STRAT}\) is
  protecting \(M\) .
  including \(\mathit{STRAT}\sb{1}\) .   ...    including \(\mathit{STRAT}\sb{j}\) .
  strat \(\mathit{sid}\sb{1}\) : \(T\sb{11}\) ... \(T\sb{1m}\) @ \(K\sb{1}\) .
  sd \(\mathit{sid}\sb{1}\)(\(\rho\sb{11}\),...,\(\rho\sb{1m}\)) := \(\sigma\sb{1}\) .    
     ...    
  strat \(\mathit{sid}\sb{n}\) : \(T\sb{n1}\) ... \(T\sb{np}\) @ \(K\sb{n}\) .
  sd \(\mathit{sid}\sb{n}\)(\(\rho\sb{n1}\),...,\(\rho\sb{np}\)) := \(\sigma\sb{n}\) .
  csd \(\mathit{sid}\sb{n}\)(\(\rho'\sb{n1}\),...,\(\rho'\sb{np}\)) := \(\sigma'\sb{n}\) if \(C\) .
endsm
\end{alltt}
}

\noindent where $M$ is the system module whose rewrites are being
controlled, $\mathit{STRAT}_1, \ldots, \mathit{STRAT}_j$ are imported strategy
submodules, $\mathit{sid}_1, \ldots, \mathit{sid}_n$ are identifiers, and
$\sigma_1, \ldots, \sigma_n$ are strategy
expressions (over the language of labels provided by $M$),
where the identifiers can appear, thus allowing (mutually) recursive definitions. 
%
%
A strategy identifier can have \emph{data arguments}, that are terms
built with the syntax defined in the system module \texttt{M}. When a
strategy identifier is declared (with the keyword \texttt{strat}), the
types of its arguments (if any) are specified between the symbols
\texttt{:} and \texttt{@}.  After the symbol \texttt{@}, the type of the
terms to which this strategy can be applied is also specified.
A strategy definition (introduced with the keyword
\texttt{sd}) associates a strategy expression (on the righthand side
of the symbol \texttt{:=}) with a strategy identifier (on the lefthand side)
with patterns as arguments, used to capture the values passed when
the strategy is invoked. 
These strategy definitions can be conditional (with keyword \texttt{csd}). 
%
A strategy module can be \emph{parametric} (see \cite{strategies/EkerMMV06} for details).

\subsection{Example}
\label{ex:river-cross-strat}

Here we show how to control by means of strategies the rewriting in module
\verb"RIVER-CROSSING" from Section~\ref{ex:river-cross}.
In \cite{maude/2007} the first two rules were presented as equations in order
to force Maude to apply them before any other rule if it is possible. But 
besides the fact that that solution introduced a coherence problem that had
to be solved, it changed the semantics of the problem. Here we can guarantee
the priority of these two rules by means of strategies. The \verb"eating"
strategy below performs all possible eatings; the \verb"oneCrossing" strategy
applies one of the other rules once; and the \verb"allCE" strategy
returns all the possible reachable states where eating has had the higher
priority. That is, \texttt{allCE} ensures that when someone can eat, it eats for sure; 
we cannot recover from a disaster situation. Finally, if the strategy 
\verb"solve" applied
to the initial state (where we assume that all the objects are located on the left riverbank) 
returns a solution, it means that there is a safe way in which the
shepherd can transport all his belongings to the other side of the river.

{\codesize
\begin{verbatim}
smod RIVER-CROSSING-STRAT is
  protecting RIVER-CROSSING .

  strat eating : @ Group . 
  sd eating := (wolf-eats | goat-eats) ! .
  
  strat oneCrossing : @ Group .
  sd oneCrossing := shepherd-alone | wolf | goat | cabbage .
  
  strat allCE : @ Group .
  sd allCE := (eating ; oneCrossing) * .
  
  strat solve : @ Group .
  sd solve := allCE ; match (s(right) w(right) g(right) c(right)) .
endsm
\end{verbatim}
}

\subsection{Implementations of the strategy language}

Using the metalevel mechanisms provided by the Maude system as a
\emph{metalanguage}, we have implemented a prototype of the Maude
strategy language \cite{MartiOlietMeseguerVerdejo04}.
The metalevel features of Maude allow the definition of operations to
work with modules and computations as objects, as is the case with
strategies.
The prototype works internally with a labelled version of the computation tree
obtained by applying a strategy to a term.
The prototype is completed with a user interface providing commands to load
modules, execute strategy expressions on states, and show results. 
The prototype and some examples can be obtained from \url{http://maude.sip.ucm.es/strategies}.

After validating the language experimentally and reaching a
 mature language design,
a direct implementation of our strategy language at
the C++ level, at which the Maude system itself is implemented, 
is currently being developed \cite{strategies/EkerMMV06}.
This will make
the language a stable new feature of Maude and will allow a more
efficient execution. 

In the meantime, we have advanced on the semantic foundations of the Maude strategy 
language in \cite{entcs/Marti-OlietMV09}. We have shown that 
a strategy language $\mathcal{S}$ can be seen as a rewrite theory transformation
$\mathcal{R} \mapsto \mathcal{S}(\mathcal{R})$ 
such that $\mathcal{S}(\mathcal{R})$ provides a way of executing $\mathcal{R}$ 
in a controlled way. One such theory transformation for the Maude strategy 
language is presented in detail in \cite{entcs/Marti-OlietMV09}, providing in this
way a rewriting semantics for the strategy language; since this rewriting semantics
is executable, we obtained a different metalevel implementation. 
We also studied in \cite{entcs/Marti-OlietMV09} some general requirements for strategy
languages. Some of these requirements, like \emph{soundness} and 
\emph{completeness} with
respect to the rewrites in $\mathcal{R}$, are absolute requirements that every
strategy language should fulfill. Other more optional requirements, that
we call \emph{monotonicity} and \emph{persistence}, 
represent the fact that no solution is
ever lost. A future research direction is the
\emph{increased performance of strategy evaluations through parallelism}.  
The point is that in $\mathcal{S}(\mathcal{R})$ a term $\sigma \at t$ (where 
strategy $\sigma$ is being applied to term $t$)
incrementally evaluates to a (possibly nested) \emph{set data structure}, 
so that the natural concurrency of rewriting logic is directly exploitable
in $\mathcal{S}(\mathcal{R})$ by applying different rules in different places 
of this data structure where solutions are generated.  This naturally suggests 
a distributed implementation of strategy languages.

\section{Some applications}
\label{sec:strat-apps}

We briefly present in this section several research areas where Maude's
strategy language has been applied successfully. 

\paragraph{Operational semantics}

Rewriting logic and Maude are a very well-known semantic framework 
\cite{Marti-OlietMeseguer02b,tcs/MeseguerR07}. By using strategies, the
semantics representations can be made more simple and powerful, by separating 
the representation of the semantic rules from the mechanism used to control
how they have to be applied. A simple example with Milner's CCS semantics is
illustrated in \cite{MartiOlietMeseguerVerdejo04}.

In the \emph{ambient calculus} \cite{cardelli98mobile} an \emph{ambient} is a 
place limited by a boundary where computations take place. Its contents are 
a parallel composition of sequential processes and subambients; communication 
between processes is local, through a blackboard. The operational semantics for 
the ambient calculus consists of a set of structural 
congruence rules and a set of reduction rules, which can be represented in Maude
by using strategies, as detailed in \cite{RosaSeguraVerdejo05}.
It gives us some congruence rules for free (as the commutativity 
and associativity of some operators), and the rest of the congruence rules are 
implemented as Maude equations.
The reduction rules are then represented as rewrite rules in Maude.
However, the reduction relation of the calculus is not a congruence for all the operators. 
This means that we cannot freely use the rewrite rules, as Maude would apply them 
anywhere in a term; and we do not want them to be applied after some operators. 
This is one of the reasons why the definition of a strategy that controls the 
application of these rules is necessary.

\emph{Eden} \cite{lop04} is a parallel extension of the functional language Haskell.
On behalf of parallelism, Eden overrides Haskell's pure lazy approach, combining
a non-strict functional application with eager process creation and communication.
The operational semantics of Eden \cite{HidalgoOrtega02} is defined by means of a 
two-level transition system: the lower level handles local effects within processes, whereas
the upper level describes the effects global to the whole system,
like process creation and data communication. Then, the global evolution of the system   
is described by iteration of the scheduling rules and the sequential composition of
other transitions. Thus the definition of the semantics itself imposes an order in the application
of the semantic rules and the use of strategies is mandatory. The representation
in Maude all of these rules and relations is studied in \cite{wrs/Hidalgo-HerreroVO07}.
Most of the semantic rules are represented as rewrite rules and the transition relations that
are defined as the concatenation or repetition of other relations are defined by
means of strategies. But there are a few semantic rules that are more abstract in their mathematical
formulation so they cannot be directly translated to rewrite rules. They are
represented as (usually recursive) strategies that combine other strategies
or rewrite rules. All of Eden's operational semantics has been represented at a quite
abstract level, independently from factors such as the eagerness degree in
the creation of new processes or the speculation degree. It was used to analyze 
algorithmic skeletons implemented in Eden in \cite{parco/Hidalgo-HerreroOR05}.

Modular structural operational semantics (MSOS) is a framework that allows structural 
operational semantics specifications to be made modular in the sense of not imposing 
the redefinition of transition rules when an extension is made. MSOS can be implemented 
in Maude in a quite precise way \cite{amast/BragaHMM00,amast/MeseguerB04}. 
The Maude MSOS Tool \cite{entcs/ChalubB07}, an executable environment for modular 
structural operational semantics, has been endowed with the possibility of defining 
strategies over its transition rules, by combining the Maude MSOS Tool with the 
Maude strategy language in \cite{sos/BragaV06}. One advantage of this combination 
is the possibility of executing Ordered SOS specifications \cite{ulidowski}, including negative premises.

\paragraph{Membrane systems}

A membrane consists of a multiset $w$ of objects, a set $R$ of evolution rules (which are ordered by a priority relation), and a control mechanism $C$ describing the way in which the rules are used to modify the multiset $w$ in an evolution step. Control mechanisms can be given by maximally parallel rewriting, maximally parallel rewriting with priorities, and maximally parallel rewriting with promoters and/or inhibitors. 
In \cite{membrane/AndreiCL06} it is shown how Maude can be used to specify membrane systems and how the control mechanisms in membranes can be described by using strategies. The strategy-based rewrite semantics thus defined preserves the maximal concurrency expressed by the maximal parallel application of the evolution rules \cite{entcs/Lucanu09}. This framework has been improved with the notion of \emph{strategy controllers} \cite{entcs/AndreiL09}, which allow to reason at the higher level of computation given by the evolution of the membrane systems. The intuition behind a strategy controller is that it decides which strategy is applied in the current state. A prototype has been developed \cite{entcs/AndreiL09} by extending the implementation at the metalevel of Maude's strategy language \cite{entcs/Marti-OlietMV09}.

\paragraph{Multi-agent systems}

One of the challenges in the design and development of multi-agent systems is to coordinate and control the behavior of individual agents. There are different approaches, from low level ones (e.g., channel-based coordination) to high level ones (e.g., normative artifacts). These normative artifacts observe the actions performed by individual agents, determine their effects in the environment (which is shared by all individual agents), determine the violations caused by performing the actions, and possibly, impose sanctions. In \cite{jucs/AstefanoaeiDMB09,iat/AstefanoaeiBD09}, the semantics of norm-based organization artifacts is specified by using Maude and its strategy language. Strategies are used as an alternative way to implement different normative artifacts without changing the semantics of the normative language. Thus, the normative multi-agent system is executed with respect to the transition rules which give the semantics of the normative language. However, how the system changes is described at an upper level by strategically instrumenting the transition rules. By using strategies there is a clear separation between executions and control. Timed choreographies are also considered in \cite{atal/AstefanoaeiBD10}.

\paragraph{Other applications}

Maude's strategy language has also been applied to solve sudoku puzzles
in \cite{entcs/Santos-GarciaP07}; to formalize web services composition
in \cite{iswsa/MerouaniMS10}; to execute a rewriting logic representation of neural networks and the 
backpropagation learning algorithm in \cite{dcai/Santos-GarciaPV08}; 
to present a rule-based approach for the design of dynamic software architectures in 
\cite{entcs/BruniLM09};
and to develop a specification of the connection method
(a goal-directed proof procedure that requires a careful control over clause copies) 
for first-order logic in \cite{entcs/HolenJW09}. Of course, there are surely more examples we are not aware of yet.

\section{Completion}
\label{sec:completion-strats}

A \emph{completion procedure} is a method used in equational logic to build from a set of (unordered) identities $E$ an equivalent canonical set of rewrite rules $R$, i.e.\ a confluent, noetherian and interreduced set of rules used to compute normal forms. A basic completion procedure can be given by firstly orienting the identities in $E$ (using a provided reduction order on terms) and then iteratively computing all critical pairs of the rewrite system obtained so far and adding to it oriented versions of all the non-joinable ones. In order to avoid the huge number of rules that this basic procedure usually generates, rules can be simplified by reducing them with the help of other rules. Following Bachmair and Dershowitz \cite{jacm/BachmairD94}, that method can be described by a set of inference rules (Figure~\ref{completion-rules}) that covers a wide range of different specific completion procedures. A specific completion procedure is obtained from that set of rules by fixing a \emph{strategy} for rule application.

\begin{figure}
\begin{center}$\renewcommand{\arraystretch}{3}
\begin{array}{l@{\hspace{3em}}c@{\hspace{1.5em}}l}
\textsc{Deduce} & \bigfrac{E, R}{E \cup \{ s \approx t\}, R} & \textrm{if } s \leftarrow_R u \rightarrow_R t \\
\textsc{Orient} & \bigfrac{E \cup  \{ s \approx t\}, R}{E, R\cup \{ s \rightarrow t\}} & \textrm{if } s > t  \\
\textsc{Delete} & \bigfrac{E\cup \{ s \approx s\}, R}{E, R} &  \\
\textsc{Simplify-identity} & \bigfrac{E\cup  \{ s \approx t\}, R}{E \cup \{ u \approx t\}, R} & \textrm{if } s \rightarrow_R u \\
\textsc{R-Simplify-rule} & \bigfrac{E, R\cup  \{ s \rightarrow t\}}{E, R\cup \{ s \rightarrow u\}} & \textrm{if } t \rightarrow_R u \\
\textsc{L-Simplify-rule} & \bigfrac{E, R\cup  \{ s \rightarrow t\}}{E\cup \{ u \approx t\}, R} & \textrm{if } s \stackrel{\sqsupset}{\rightarrow}_R u \\
\end{array}$
\end{center}
\caption{Inference rules for completion.}\label{completion-rules}
\end{figure}

The inference rules work on pairs $(E,R)$ where $E$ is a finite set of identities (input identities or critical pairs that have not yet been transformed into rules) and $R$ is a finite set of terminating rewrite rules.\footnote{Termination of $R$ is ensured by a reduction order $>$ that is given as an input to the completion procedure.} The goal is to transform an initial pair $(E_0,\emptyset)$ into a pair $(\emptyset,R)$ such that $R$ is convergent and equivalent to $E_0$.

The inference rule \textsc{Deduce} derives an identity that is a direct consequence of rules in $R$, and adds it to $E$. A special, common case is adding a critical pair of $R$ to $E$. The rule \textsc{Orient} takes an identity that can be ordered with the help of $>$ and adds the corresponding rule to $R$. The rule \textsc{Delete} removes a trivial identity, and the rule \textsc{Simplify-identity} uses $R$ to reduce identities. Both rules can be used together to remove joinable critical pairs. The rule \textsc{R-Simplify-rule} reduces the righthand side of a rule. Since, by assumption, termination of $R$ can be shown using $>$, we know that $s \rightarrow_R t \rightarrow_R u$ implies $s > t > u$. For this reason, $s \rightarrow u$ can be kept as a rule. However, when reducing the lefthand side of a rule $s \rightarrow t$ to $u$, it is not clear whether $u > t$ is satisfied. Thus, the rule \textsc{L-Simplify-rule} adds $u \approx t$ as an identity. The notation $s \stackrel{\sqsupset}{\rightarrow}_R u$ is used to express that $s$ is reduced by a rule $l \rightarrow r \in R$ such that $l$ cannot be reduced by $s \rightarrow t$.

As mentioned above, specific completion procedures can be obtained from the inference rules in Figure~\ref{completion-rules} by fixing a strategy for rule application. In the following sections we show how to use Maude's strategy language to represent several completion procedures described by Lescanne in \cite{Lescanne89}. ELAN \cite{BorovanskyKirchnerKirchnerMoreau02} has also been used to prototype completion algorithms in \cite{rta/KirchnerM95} by using strategies and constraints. Even Maude has already been used to represent Huet's completion algorithm \cite{jcss/Huet81}, by using strategies defined at the metalevel in \cite{Clavel-Meseguer97}.

In \cite{Lescanne89} several algorithms for completion are presented using the functional programming language CAML. The main differences among the algorithms are the data structures on which the transition rules operate, and
the control that describes the way the transitions rules are invoked.
Here, we redo this work in Maude by using the proposed strategy language:
the data structure is the term being rewritten,
inference rules (transition rules in Lescanne's terminology) are represented as rewrite rules, and 
the control is represented as strategies applying the rules in a directed way.
We consider three algorithms, as described by Lescanne: N-Completion, S-Completion, and ANS-Completion. Each algorithm is a refinement of the previous one, obtained by
adding more components to the data structure,
adapting the transition rules, and 
changing the control with the idea of making the algorithm more efficient. 
In this presentation we do not deal with the two unfailing completion algorithms also
presented by Lescanne. 


\subsection{N-completion}\label{Ncompletion}

N-completion is a first improvement of the overly abstract inference rules in order to take the computation of critical pairs into account. For this algorithm, the data structure has three components:
\begin{itemize}
\item $E$ is a set of identities, either given identities or computed critical pairs,\footnote{The
fact that an identity is an unordered pair is represented in Maude by using an operator \texttt{\char`_=.\char`_} with
the commutative attribute.}
\item $T$ is a set of rules whose critical pairs have not been computed yet, and
\item $R$ is another set of rules whose critical pairs have already been computed
 (marked rules in Huet's terminology \cite{jcss/Huet81}).
\end{itemize}

The transition rules correspond to the inference rules but adapted to these three components. They are represented in Maude as follows:

{\codesize
\begin{verbatim}
mod N-COMPLETION is  
  pr CRITICAL-PAIRS .
  
  sort System .
  op <_,_,_> : RlS RlS EqS -> System .  *** < R, T, E >
  
  var E : EqS .  var r : Rl .  vars R T : RlS .  vars s t u : Term .
  
  rl [Deduce] : < R, T, E > => < R, T, E s =. t > . *** if  s <-- u --> t
    
  crl [Orient] : < R, T, E s =. t > => < R, T s -> t, E >  if s > t .
  
  rl [Delete] : < R, T, E s =. s > => < R, T, E > .
  
  crl [Simplify] : < R, T, E s =. t > => < R, T, E u =. t >
   if u := reduce(s, R T) . 
   
  crl [R-Simplify] : < R s -> t, T, E > => < R s -> u, T, E >
   if u := reduce(t, R T) . 
   
  crl [R-Simplify] : < R, T s -> t, E > => < R, T s -> u, E >
   if u := reduce(t, R T) . 
   
  crl [L-Simplify] : < R s -> t, T, E > => < R, T, E u =. t >
   if u := reduce>(s -> t, R T) . 
   
  crl [L-Simplify] : < R, T s -> t, E > => < R, T, E u =. t >
   if u := reduce>(s -> t, R T) . 
   
  rl [move] : < R, r T, E > => < r R, T, E > .
endm
\end{verbatim}
}

The included module \verb"CRITICAL-PAIRS" contains functions, defined at the metalevel, that compute critical pairs of a set of rules, or the reductions $\rightarrow_R$ (\verb"reduce") and $\stackrel{\sqsupset}{\rightarrow}_R$ (\verb"reduce>"). Observe that the inference rule \textsc{R-Simplify-rule} gives rise to two rewrite rules: one where the righthand side of a rule in the set $R$ is simplified and another one where the righthand side of a rule in $T$ is simplified. The same happens with \textsc{L-Simplify-rule}. The rewrite rule \verb"move" will be used by the strategy to move a rule from set $T$ to set $R$.\footnote{That is because the only way a strategy can modify the term being rewritten is by means of rewrite rules.}

The algorithm N-completion has essentially three steps, namely \emph{success}, when $T$ and $E$ are empty, \emph{computing critical pairs} after simplification of the rules, when $E$ is empty, and \emph{orienting} an identity into a rule after simplification of the identities, when $E$ is not empty. In the orientation part, it could happen that by simplification all the identities disappear. In this case, one does nothing, that is just translated by a recursive call. This algorithm can be succinctly expressed by the following strategy module (the strategy identifier declarations have been omitted):

{\codesize
\begin{verbatim}
smod N-COMPLETION-STRAT is
  protecting N-COMPLETION .
  
  sd N-COMP :=  success orelse deduce orelse orient .
  
  sd success := match < R, mtRlS, mtEqS > .
  
  sd deduce := match < R, r T, mtEqS > ;
               deduction ;
               simplify-rules ;
               N-COMP .
  
  sd deduction := matchrew < R, r' T, E > by
                    < R, r' T, E > using (add-crit-pairs(CP(r', R r')) ;
                                          move[r <- r']) .
                   
  sd add-crit-pairs(mtEqS) := idle .
  sd add-crit-pairs(s1 =. t1 E) := Deduce[s <- s1 ; t <- t1] ; add-crit-pairs(E) .
  
  sd simplify-rules := (L-Simplify | R-Simplify) ! .
  
  sd orient := match < R, T, e E >  ;
               simplify-eqs ;
               ( (match < R, T, mtEqS > ; N-COMP)
                 orelse (Orient ; N-COMP) ) .
  
  sd simplify-eqs := (Delete | Simplify) ! .
endsm
\end{verbatim}
}

\subsubsection{Example}

Let us consider the following set of equations
\[ \{\; g(x,y) \approx a, g(x,y) \approx h(x,y), h(x,y) \approx f(x), h(x,y) \approx f(y) \;\} \]
and the lexicographic path order induced by the precedence $g > h > f > a$.
The basic completion procedure uses \textsc{Orient} to generate the rules
\[ \{\; g(x,y) \rightarrow a, g(x,y) \rightarrow h(x,y), h(x,y) \rightarrow f(x), h(x,y) \rightarrow f(y) \;\} \]
and then \textsc{Deduce} to compute the critical pairs $a\approx h(x,y)$ and $f(x)\approx f(y)$. We can reproduce this behavior by using some of the previous strategies

{\codesize
\begin{verbatim}
Maude> (srew < mtRlS, mtRlS, eqs > using Orient ! .)
result System :
    < mtRlS, 'g['x:S,'y:S] -> 'a.S     'g['x:S,'y:S] -> 'h['x:S,'y:S]
             'h['x:S,'y:S] -> 'f['x:S] 'h['x:S,'y:S] -> 'f['y:S], mtEqS >

Maude> (cont using deduction ! ; Delete ! .)
result System :
    < 'g['x:S,'y:S] -> 'a.S      'g['x:S,'y:S] -> 'h['x:S,'y:S]
      'h['x:S,'y:S] -> 'f['x:S]  'h['x:S,'y:S] -> 'f['y:S],
    mtRlS, 'a.S =. 'h['x1:S,'y1:S]  'f['x1:S] =. 'f['y1:S] >
\end{verbatim}
}

\noindent where \verb"srew" is the command for applying a strategy to a given term; the constant \verb"eqs" represents the above set of equations; \verb"cont" is the command used to apply a strategy to the term returned by the previous application; and all the terms are metarepresented.

The basic completion procedure continues by trying to simplify and orient these critical pairs, but the terms in $f(x)\approx f(y)$ are irreducible and cannot be compared with any reduction order. So, the procedure fails. However, $a\approx h(x,y)$ can be oriented and used to compute new critical pairs that allow to reduce $f(x)\approx f(y)$ to a trivial identity. The N-completion procedure is able to find this solution.

{\codesize
\begin{verbatim}
Maude> (srew < mtRlS, mtRlS, eqs > using N-COMP .)
result System :
    < 'f['x:S] -> 'a.S  'g['x:S,'y:S] -> 'a.S  'h['x:S,'y:S] -> 'a.S, mtRlS, mtEqS >
\end{verbatim}
}

\subsection{S-completion}

The main aim of orienting identities is to use them to simplify whenever it is possible. However, N-completion makes a bad use of simplification. S-completion is an improvement of N-completion where a rule is used for simplification as soon as it has been generated. When an identity is oriented into a rule, it enters a set $S$ where it is used to simplify all the other identities and rules. Thus, the data structure has now four components:
\begin{itemize}
\item $E$ is a set of identities, like in N-completion,
\item $S$ is a (singleton or empty) set of oriented identities (rules) 
that are used to simplify other rules,
\item $T$ is a set of rules already used for simplifying, 
but whose critical pairs have not been computed yet, and
\item $R$ is another  set of rules whose critical pairs have already been computed,
like in N-completion.
\end{itemize}

The only difference with the N-completion is the set $S$ through which a rule has to go before entering $T$. The rewrite rules are modified to express this fact. We only show the rule that really changes (\verb"Orient") and a new rule that is used by the strategy to join the sets $T$ and $S$. The rest of the rules are only modified by including the new set $S$.

{\codesize
\begin{verbatim}
mod S-COMPLETION is  
  pr CRITICAL-PAIRS .  
  sort System .
  op <_,_,_,_> : RlS RlS RlS EqS -> System .  *** < R, T, S, E >
  
  crl [Orient] : < R, T, S, E s =. t > => < R, T, S s -> t, E >  if s > t .
    
  rl [concatT&S] : < R, T, S, E > => < R, T S, mtRlS, E > .
  
  [...]
endm
\end{verbatim}
}

The simplification step is clearly distinguished from the three others. It is performed when $S$ is not empty. The completion process ends when there are no more identities or rules in $E$, $S$ or $T$. Again, we only show the strategies that really change.

{\codesize
\begin{verbatim}
smod S-COMPLETION-STRAT is
  
  sd S-COMP :=  success orelse simplify-rules orelse deduce orelse orient .
  
  sd success := match < R, mtRlS, mtRlS, mtEqS > .
  
  sd simplify-rules := match < R, T, r S, E > ;
                       (L-Simplify | R-Simplify) ! ;
                       concatT&S ;
                       S-COMP .
  
  sd deduce := match < R, r T, mtRlS, mtEqS > ;
               deduction ;
               S-COMP .
  [...]
endsm 
\end{verbatim}
}

%
%
%

\subsection{ANS-completion}

S-completion computes all the critical pairs between all rules in $R$ and 
one rule in $T$. In order to apply simplification as soon as possible, it is better to compute the 
critical pairs between one rule in $R$ and one rule in $T$ at a time. A new set $C$ is created to contain one rule extracted from $T$ with which critical pairs with rules in $R$ are computed. To keep track of the rules whose critical pairs are computed with the rule in $C$, $R$ is split into two sets, $A$ (for already computed) and $N$ (for not yet computed). The data structure has now six components:
\begin{itemize}
\item $E$ is a set of identities, like in S-completion,
\item $S$ is a set of simplifying rules, like in S-completion,
\item $T$ is a set of rules coming from $S$ and waiting to enter $C$,
\item $C$ is a set that contains at most one rule and whose critical pairs are computed 
with one in $N$,
\item $N$ is the part of $R$ whose critical pairs have not been computed with $C$ but 
whose critical pairs with $A \cup N$ have been computed, and
\item $A$ is a set whose critical pairs with $A \cup N \cup C$ have been computed.
\end{itemize}

The transition rules are trivially adapted to work with this new data structure, and two new rules are added. Rule \verb"AC2N" joins the sets $A$ and $C$ with $N$ and rule \verb"fillC" extracts the smallest rule in $T$ and puts it in $C$.

{\codesize
\begin{verbatim}
mod ANS-COMPLETION is  
  pr CRITICAL-PAIRS .
  
  sort System .
  op <_,_,_,_,_,_> : RlS RlS RlS RlS RlS EqS -> System .  *** < A, N, C, T, S, E >
  
  [...]
  
  rl [AC2N] : < A, N, C, T, S, E > => < mtRlS, A N C, mtRlS, T, S, E > .
  
  crl [fillC] : < A, N, C, T, S, E > => < mtRlS, A N, r, T', S, E > 
            if  r := least-rule(T) /\
                r T' := T .
endm
\end{verbatim}
}

The procedure has now six parts, namely \emph{success}, \emph{simplification}, \emph{orientation}, \emph{deduction}, \emph{internal deduction}, and \emph{beginning of a new loop} of computation of critical pairs. \emph{Deduction} computes the critical pairs between the smallest rule in $N$ and the rule in $C$, whereas \emph{internal deduction} computes the critical pairs obtained by superposing the rule in $C$ on itself.

{\codesize
\begin{verbatim}
smod ANS-COMPLETION-STRAT is

  sd ANS-COMP :=  success             orelse 
                  simplify-rules      orelse 
                  orient              orelse
                  deduce              orelse 
                  internal-deduction  orelse
                  new-loop .
  
  sd success := match ( < A, N, mtRlS, mtRlS, mtRlS, mtEqS > ) .
  
  sd deduce := match (< A, r N, r', T, mtRlS, mtEqS >) ;
               deduction ;
               ANS-COMP .
  
  sd deduction := matchrew < A, r' N, r'', T, S, E > s.t. r' := least-rule(r' N) by
                    < A, r' N, r'', T, S, E > using (add-crit-pairs(CP(r'', r')) ;
                                                     move[r <- r']) .
                   
  sd internal-deduction := ( matchrew < A, mtRlS, r', T, mtRlS, mtEqS > by
                < A, mtRlS, r', T, mtRlS, mtEqS > using (add-crit-pairs(CP(r', r')) ;
                                                         AC2N) ) ;
                            ANS-COMP .
   
  sd new-loop := fillC ; ANS-COMP .
  [...]
endsm
\end{verbatim}
}

\section{Concluding remarks}

As another application of the Maude strategy language, we have described
completion procedures as transition rules + control, following Lescanne's
proposal \cite{Lescanne89}.  Our version, using rewrite rules and declarative
strategies instead of CAML programs, is more abstract and emphasizes the fact 
that inference rules do not change at all in the different algorithms. 

This new case study on top of a growing list of applications throughout the world,
has confirmed that the design of the Maude strategy language 
is good enough to handle a variety of algorithms based on the 
transition rules + control paradigm, using Lescanne's words, or simply
data + actions + strategies, using the well-known three-level approach to 
problem solving.

As part of ongoing work, we are studying the integration of the strategies and
model-checking features of Maude.
Properties of rewriting systems expressed in linear temporal logic
can be studied in Maude with the help of
  its integrated model checker \cite{spin/EkerMS03,maude/2007}. 
  This tool checks the temporal properties
  fulfilled by a transition system by considering all the possible executions
  from a given state; however, as described in this paper, we can be interested in
  using the Maude strategy language to control those executions and possibly to
  restrict them, thus modifying the transition system. Thus, we plan to study how to
  model check temporal formulas satisfied by systems controlled by strategies,
  thus combining the advantages of two quite useful Maude features for the specification
  of systems: model checking at the property level and strategies at the
  execution level. 

\subsubsection*{Acknowledgments} 

We are very grateful to the WRS 2011 organizers, and in particular to Santiago
Escobar, for the opportunity to present
this work in such an agreeable environment; and to our colleagues for their
collaboration in all the different aspects of the research reported in the first part
of this paper.


\begin{thebibliography}{10}
\providecommand{\bibitemdeclare}[2]{}
\providecommand{\urlprefix}{Available at }
\providecommand{\url}[1]{\texttt{#1}}
\providecommand{\href}[2]{\texttt{#2}}
\providecommand{\urlalt}[2]{\href{#1}{#2}}
\providecommand{\doi}[1]{doi:\urlalt{http://dx.doi.org/#1}{#1}}
\providecommand{\bibinfo}[2]{#2}

\bibitemdeclare{inproceedings}{membrane/AndreiCL06}
\bibitem{membrane/AndreiCL06}
\bibinfo{author}{Oana Andrei}, \bibinfo{author}{Gabriel Ciobanu} \&
  \bibinfo{author}{Dorel Lucanu} (\bibinfo{year}{2006}):
  \emph{\bibinfo{title}{Expressing Control Mechanisms of Membranes by Rewriting
  Strategies}}.
\newblock In \bibinfo{editor}{Hendrik~Jan Hoogeboom}, \bibinfo{editor}{Gheorghe
  Paun}, \bibinfo{editor}{Grzegorz Rozenberg} \& \bibinfo{editor}{Arto
  Salomaa}, editors: {\sl \bibinfo{booktitle}{Membrane Computing, 7th
  International Workshop, WMC 2006, Leiden, The Netherlands, July 17-21, 2006,
  Revised, Selected, and Invited Papers}}, {\sl \bibinfo{series}{Lecture Notes
  in Computer Science}} \bibinfo{volume}{4361}, \bibinfo{publisher}{Springer},
  pp. \bibinfo{pages}{154--169}.
\newblock \urlprefix\url{http://dx.doi.org/10.1007/11963516_10}.

\bibitemdeclare{inproceedings}{entcs/AndreiL09}
\bibitem{entcs/AndreiL09}
\bibinfo{author}{Oana Andrei} \& \bibinfo{author}{Dorel Lucanu}
  (\bibinfo{year}{2009}): \emph{\bibinfo{title}{Strategy-Based Proof Calculus
  for Membrane Systems}}.
\newblock In \bibinfo{editor}{Ro{\c{s}}u}  \cite{wrla/2008}, pp.
  \bibinfo{pages}{23--43}.
\newblock \urlprefix\url{http://dx.doi.org/10.1016/j.entcs.2009.05.011}.

\bibitemdeclare{inproceedings}{iat/AstefanoaeiBD09}
\bibitem{iat/AstefanoaeiBD09}
\bibinfo{author}{Lacramioara Astefanoaei}, \bibinfo{author}{Frank~S. de~Boer}
  \& \bibinfo{author}{Mehdi Dastani} (\bibinfo{year}{2009}):
  \emph{\bibinfo{title}{Rewriting Agent Societies Strategically}}.
\newblock In: {\sl \bibinfo{booktitle}{Proceedings of the 2009 IEEE/WIC/ACM
  International Conference on Web Intelligence and International Conference on
  Intelligent Agent Technology - Workshops, Milan, Italy, September 15-18,
  2009}}, \bibinfo{publisher}{IEEE}, pp. \bibinfo{pages}{441--444}.
\newblock \urlprefix\url{http://dx.doi.org/10.1109/WI-IAT.2009.321}.

\bibitemdeclare{inproceedings}{atal/AstefanoaeiBD10}
\bibitem{atal/AstefanoaeiBD10}
\bibinfo{author}{Lacramioara Astefanoaei}, \bibinfo{author}{Frank~S. de~Boer}
  \& \bibinfo{author}{Mehdi Dastani} (\bibinfo{year}{2010}):
  \emph{\bibinfo{title}{Strategic executions of choreographed timed normative
  multi-agent systems}}.
\newblock In \bibinfo{editor}{Wiebe van~der Hoek}, \bibinfo{editor}{Gal~A.
  Kaminka}, \bibinfo{editor}{Yves Lesp{\'e}rance}, \bibinfo{editor}{Michael
  Luck} \& \bibinfo{editor}{Sandip Sen}, editors: {\sl
  \bibinfo{booktitle}{Proceedings of the 9th International Conference on
  Autonomous Agents and Multiagent Systems, AAMAS 2010, Toronto, Canada, May
  10-14, 2010, Volume 1-3}}, \bibinfo{publisher}{IFAAMAS}, pp.
  \bibinfo{pages}{965--972}.
\newblock \urlprefix\url{http://doi.acm.org/10.1145/1838206.1838336}.

\bibitemdeclare{article}{jucs/AstefanoaeiDMB09}
\bibitem{jucs/AstefanoaeiDMB09}
\bibinfo{author}{Lacramioara Astefanoaei}, \bibinfo{author}{Mehdi Dastani},
  \bibinfo{author}{John-Jules~Ch. Meyer} \& \bibinfo{author}{Frank~S. de~Boer}
  (\bibinfo{year}{2009}): \emph{\bibinfo{title}{On the Semantics and
  Verification of Normative Multi-Agent Systems}}.
\newblock {\sl \bibinfo{journal}{Journal of Universal Computer Science}}
  \bibinfo{volume}{15}(\bibinfo{number}{13}), pp. \bibinfo{pages}{2629--2652}.
\newblock \urlprefix\url{http://www.jucs.org/jucs_15_13/on_the_semantics_and}.

\bibitemdeclare{article}{jacm/BachmairD94}
\bibitem{jacm/BachmairD94}
\bibinfo{author}{Leo Bachmair} \& \bibinfo{author}{Nachum Dershowitz}
  (\bibinfo{year}{1994}): \emph{\bibinfo{title}{Equational Inference, Canonical
  Proofs, and Proof Orderings}}.
\newblock {\sl \bibinfo{journal}{Journal of the ACM}}
  \bibinfo{volume}{41}(\bibinfo{number}{2}), pp. \bibinfo{pages}{236--276}.
\newblock \urlprefix\url{http://doi.acm.org/10.1145/174652.174655}.

\bibitemdeclare{manual}{TOM-manual}
\bibitem{TOM-manual}
\bibinfo{author}{Emilie Balland}, \bibinfo{author}{Paul Brauner},
  \bibinfo{author}{Radu Kopetz}, \bibinfo{author}{Pierre-Etienne Moreau} \&
  \bibinfo{author}{Antoine Reilles} (\bibinfo{year}{2006}):
  \emph{\bibinfo{title}{{TOM} Manual}}.
\newblock \urlprefix\url{http://tom.loria.fr}.

\bibitemdeclare{article}{BorovanskyKirchnerKirchnerMoreau02}
\bibitem{BorovanskyKirchnerKirchnerMoreau02}
\bibinfo{author}{Peter Borovansk\'y}, \bibinfo{author}{Claude Kirchner},
  \bibinfo{author}{H{\'e}l{\`e}ne Kirchner} \& \bibinfo{author}{Pierre-Etienne
  Moreau} (\bibinfo{year}{2002}): \emph{\bibinfo{title}{{ELAN} from a rewriting
  logic point of view}}.
\newblock {\sl \bibinfo{journal}{Theoretical Computer Science}}
  \bibinfo{volume}{285}(\bibinfo{number}{2}), pp. \bibinfo{pages}{155--185}.

\bibitemdeclare{article}{BKKR01}
\bibitem{BKKR01}
\bibinfo{author}{Peter Borovansk\'y}, \bibinfo{author}{Claude Kirchner},
  \bibinfo{author}{H{\'e}l{\`e}ne Kirchner} \& \bibinfo{author}{Christophe
  Ringeissen} (\bibinfo{year}{2001}): \emph{\bibinfo{title}{Rewriting with
  Strategies in {ELAN}: A Functional Semantics}}.
\newblock {\sl \bibinfo{journal}{International Journal of Foundations of
  Computer Science}} \bibinfo{volume}{12}, pp. \bibinfo{pages}{69--95}.

\bibitemdeclare{article}{ijfcs/BorovanskyKKR01}
\bibitem{ijfcs/BorovanskyKKR01}
\bibinfo{author}{Peter Borovansk{\'y}}, \bibinfo{author}{Claude Kirchner},
  \bibinfo{author}{H{\'e}l{\`e}ne Kirchner} \& \bibinfo{author}{Christophe
  Ringeissen} (\bibinfo{year}{2001}): \emph{\bibinfo{title}{Rewriting with
  Strategies in {ELAN}: A Functional Semantics}}.
\newblock {\sl \bibinfo{journal}{International Journal of Foundations of
  Computer Science}} \bibinfo{volume}{12}(\bibinfo{number}{1}), pp.
  \bibinfo{pages}{69--95}.
\newblock \urlprefix\url{http://dx.doi.org/10.1142/S0129054101000412}.

\bibitemdeclare{article}{tcs/BouhoulaJM00}
\bibitem{tcs/BouhoulaJM00}
\bibinfo{author}{Adel Bouhoula}, \bibinfo{author}{Jean-Pierre Jouannaud} \&
  \bibinfo{author}{Jos{\'e} Meseguer} (\bibinfo{year}{2000}):
  \emph{\bibinfo{title}{Specification and proof in membership equational
  logic}}.
\newblock {\sl \bibinfo{journal}{Theoretical Computer Science}}
  \bibinfo{volume}{236}(\bibinfo{number}{1-2}), pp. \bibinfo{pages}{35--132}.
\newblock \urlprefix\url{http://dx.doi.org/10.1016/S0304-3975(99)00206-6}.

\bibitemdeclare{misc}{BPEL}
\bibitem{BPEL}
\bibinfo{author}{BPEL} (\bibinfo{year}{2007}): \emph{\bibinfo{title}{Web
  Services Business Process Execution Language ({WS-BPEL}). {Version} 2.0}}.
\newblock \bibinfo{howpublished}{{OASIS} Standard}.

\bibitemdeclare{misc}{BPMN}
\bibitem{BPMN}
\bibinfo{author}{BPMN} (\bibinfo{year}{2011}): \emph{\bibinfo{title}{Business
  Process Model and Notation (BPMN). Version 2.0}}.
\newblock \bibinfo{howpublished}{{OMG} Specification}.

\bibitemdeclare{inproceedings}{amast/BragaHMM00}
\bibitem{amast/BragaHMM00}
\bibinfo{author}{Christiano Braga}, \bibinfo{author}{Edward~Hermann Haeusler},
  \bibinfo{author}{Jos{\'e} Meseguer} \& \bibinfo{author}{Peter~D. Mosses}
  (\bibinfo{year}{2000}): \emph{\bibinfo{title}{{M}aude Action Tool: Using
  Reflection to Map Action Semantics to Rewriting Logic}}.
\newblock In \bibinfo{editor}{Teodor Rus}, editor: {\sl
  \bibinfo{booktitle}{Algebraic Methodology and Software Technology. 8th
  International Conference, AMAST 2000, Iowa City, Iowa, USA, May 20-27, 2000,
  Proceedings}}, {\sl \bibinfo{series}{Lecture Notes in Computer Science}}
  \bibinfo{volume}{1816}, \bibinfo{publisher}{Springer}, pp.
  \bibinfo{pages}{407--421}.
\newblock \urlprefix\url{http://dx.doi.org/10.1007/3-540-45499-3_29}.

\bibitemdeclare{inproceedings}{sos/BragaV06}
\bibitem{sos/BragaV06}
\bibinfo{author}{Christiano Braga} \& \bibinfo{author}{Alberto Verdejo}
  (\bibinfo{year}{2007}): \emph{\bibinfo{title}{Modular Structural Operational
  Semantics with Strategies}}.
\newblock In \bibinfo{editor}{Rob van Glabbeek} \& \bibinfo{editor}{Peter~D.
  Mosses}, editors: {\sl \bibinfo{booktitle}{Proceedings of the Third Workshop
  on Structural Operational Semantics, SOS 2006, Bonn, Germany, August 26,
  2006}}, {\sl \bibinfo{series}{Electronic Notes in Theoretical Computer
  Science}} \bibinfo{volume}{175(1)}, \bibinfo{publisher}{Elsevier}, pp.
  \bibinfo{pages}{3--17}.
\newblock \urlprefix\url{http://dx.doi.org/10.1016/j.entcs.2006.10.024}.

\bibitemdeclare{inproceedings}{entcs/BruniLM09}
\bibitem{entcs/BruniLM09}
\bibinfo{author}{Roberto Bruni}, \bibinfo{author}{Alberto Lluch-Lafuente} \&
  \bibinfo{author}{Ugo Montanari} (\bibinfo{year}{2009}):
  \emph{\bibinfo{title}{Hierarchical Design Rewriting with {M}aude}}.
\newblock In \bibinfo{editor}{Ro{\c{s}}u}  \cite{wrla/2008}, pp.
  \bibinfo{pages}{45--62}.
\newblock \urlprefix\url{http://dx.doi.org/10.1016/j.entcs.2009.05.012}.

\bibitemdeclare{inproceedings}{cardelli98mobile}
\bibitem{cardelli98mobile}
\bibinfo{author}{Luca Cardelli} \& \bibinfo{author}{Andrew~D. Gordon}
  (\bibinfo{year}{1998}): \emph{\bibinfo{title}{Mobile Ambients}}.
\newblock In \bibinfo{editor}{Maurice Nivat}, editor: {\sl
  \bibinfo{booktitle}{Foundations of Software Science and Computation
  Structures, First International Conference, FoSSaCS'98 Held as Part of the
  Joint European Conferences on Theory and Practice of Software, ETAPS'98
  Lisbon, Portugal, March 28--April 4, 1998 Proceedings}}, {\sl
  \bibinfo{series}{Lecture Notes in Computer Science}} \bibinfo{volume}{1378},
  \bibinfo{publisher}{Springer}, pp. \bibinfo{pages}{140--155}.

\bibitemdeclare{inproceedings}{entcs/ChalubB07}
\bibitem{entcs/ChalubB07}
\bibinfo{author}{Fabricio Chalub} \& \bibinfo{author}{Christiano Braga}
  (\bibinfo{year}{2007}): \emph{\bibinfo{title}{{Maude MSOS} Tool}}.
\newblock In \bibinfo{editor}{Denker} \& \bibinfo{editor}{Talcott}
  \cite{wrla/2006}, pp. \bibinfo{pages}{133--146}.
\newblock \urlprefix\url{http://dx.doi.org/10.1016/j.entcs.2007.06.012}.

\bibitemdeclare{manual}{CDELMOMT11}
\bibitem{CDELMOMT11}
\bibinfo{author}{Manuel Clavel}, \bibinfo{author}{Francisco Dur{\'a}n},
  \bibinfo{author}{Steven Eker}, \bibinfo{author}{Patrick Lincoln},
  \bibinfo{author}{Narciso Mart{\'\i}-Oliet}, \bibinfo{author}{Jos{\'e}
  Meseguer} \& \bibinfo{author}{Carolyn Talcott} (\bibinfo{year}{2011}):
  \emph{\bibinfo{title}{{Maude} Manual (Version 2.6)}}.
\newblock \urlprefix\url{http://maude.cs.uiuc.edu/maude2-manual}.

\bibitemdeclare{book}{maude/2007}
\bibitem{maude/2007}
\bibinfo{author}{Manuel Clavel}, \bibinfo{author}{Francisco Dur{\'a}n},
  \bibinfo{author}{Steven Eker}, \bibinfo{author}{Patrick Lincoln},
  \bibinfo{author}{Narciso Mart{\'\i}-Oliet}, \bibinfo{author}{Jos{\'e}
  Meseguer} \& \bibinfo{author}{Carolyn~L. Talcott} (\bibinfo{year}{2007}):
  \emph{\bibinfo{title}{All About Maude - A High-Performance Logical Framework,
  How to Specify, Program and Verify Systems in Rewriting Logic}}.
\newblock {\sl \bibinfo{series}{Lecture Notes in Computer Science}}
  \bibinfo{volume}{4350}, \bibinfo{publisher}{Springer}.
\newblock \urlprefix\url{http://dx.doi.org/10.1007/978-3-540-71999-1}.

\bibitemdeclare{inproceedings}{Clavel-Meseguer97}
\bibitem{Clavel-Meseguer97}
\bibinfo{author}{Manuel Clavel} \& \bibinfo{author}{Jos{\'e} Meseguer}
  (\bibinfo{year}{1997}): \emph{\bibinfo{title}{Internal Strategies in a
  Reflective Logic}}.
\newblock In \bibinfo{editor}{Bernhard Gramlich} \&
  \bibinfo{editor}{H{\'e}l{\`e}ne Kirchner}, editors: {\sl
  \bibinfo{booktitle}{Proceedings of the CADE-14 Workshop on Strategies in
  Automated Deduction}}, \bibinfo{address}{Townsville, Australia}, pp.
  \bibinfo{pages}{1--12}.

\bibitemdeclare{proceedings}{wrla/2006}
\bibitem{wrla/2006}
\bibinfo{editor}{Grit Denker} \& \bibinfo{editor}{Carolyn Talcott}, editors
  (\bibinfo{year}{2007}): \emph{\bibinfo{title}{Proceedings of the Sixth
  International Workshop on Rewriting Logic and its Applications, WRLA 2006,
  Vienna, Austria, April 1-2, 2006}}. {\sl \bibinfo{series}{Electronic Notes in
  Theoretical Computer Science}} \bibinfo{volume}{176(4)},
  \bibinfo{publisher}{Elsevier}.

\bibitemdeclare{inproceedings}{strategies/EkerMMV06}
\bibitem{strategies/EkerMMV06}
\bibinfo{author}{Steven Eker}, \bibinfo{author}{Narciso Mart{\'\i}-Oliet},
  \bibinfo{author}{Jos{\'e} Meseguer} \& \bibinfo{author}{Alberto Verdejo}
  (\bibinfo{year}{2007}): \emph{\bibinfo{title}{Deduction, Strategies, and
  Rewriting}}.
\newblock In \bibinfo{editor}{Myla Archer}, \bibinfo{editor}{Thierry~Boy de~la
  Tour} \& \bibinfo{editor}{C{\'e}sar Mu{\~n}oz}, editors: {\sl
  \bibinfo{booktitle}{Proceedings of the 6th International Workshop on
  Strategies in Automated Deduction, STRATEGIES 2006, Seattle, WA, USA, August
  16, 2006}}, {\sl \bibinfo{series}{Electronic Notes in Theoretical Computer
  Science}} \bibinfo{volume}{174(11)}, \bibinfo{publisher}{Elsevier}, pp.
  \bibinfo{pages}{3--25}.
\newblock \urlprefix\url{http://dx.doi.org/10.1016/j.entcs.2006.03.017}.

\bibitemdeclare{inproceedings}{spin/EkerMS03}
\bibitem{spin/EkerMS03}
\bibinfo{author}{Steven Eker}, \bibinfo{author}{Jos{\'e} Meseguer} \&
  \bibinfo{author}{Ambarish Sridharanarayanan} (\bibinfo{year}{2003}):
  \emph{\bibinfo{title}{The {Maude} {LTL} Model Checker and Its
  Implementation}}.
\newblock In \bibinfo{editor}{Thomas Ball} \& \bibinfo{editor}{Sriram~K.
  Rajamani}, editors: {\sl \bibinfo{booktitle}{Model Checking Software, 10th
  International SPIN Workshop. Portland, OR, USA, May 9-10, 2003,
  Proceedings}}, {\sl \bibinfo{series}{Lecture Notes in Computer Science}}
  \bibinfo{volume}{2648}, \bibinfo{publisher}{Springer}, pp.
  \bibinfo{pages}{230--234}.
\newblock \urlprefix\url{http://dx.doi.org/10.1007/3-540-44829-2_16}.

\bibitemdeclare{mastersthesis}{Henche10}
\bibitem{Henche10}
\bibinfo{author}{Laura Henche} (\bibinfo{year}{2009}):
  \emph{\bibinfo{title}{Introducci\'on a la notaci\'on {BPMN} y su relaci\'on
  con las estrategias del lenguaje {Maude}}}.
\newblock Master's thesis, \bibinfo{school}{Facultad de Inform\'atica,
  Universidad Complutense de Madrid, Spain}.

\bibitemdeclare{article}{HidalgoOrtega02}
\bibitem{HidalgoOrtega02}
\bibinfo{author}{Mercedes Hidalgo-Herrero} \& \bibinfo{author}{Yolanda
  Ortega-Mall\'en} (\bibinfo{year}{2002}): \emph{\bibinfo{title}{An Operational
  Semantics for the Parallel Language {Eden}}}.
\newblock {\sl \bibinfo{journal}{Parallel Processing Letters}}
  \bibinfo{volume}{12}(\bibinfo{number}{2}), pp. \bibinfo{pages}{211--228}.

\bibitemdeclare{inproceedings}{parco/Hidalgo-HerreroOR05}
\bibitem{parco/Hidalgo-HerreroOR05}
\bibinfo{author}{Mercedes Hidalgo-Herrero}, \bibinfo{author}{Yolanda
  Ortega-Mall{\'e}n} \& \bibinfo{author}{Fernando Rubio}
  (\bibinfo{year}{2005}): \emph{\bibinfo{title}{Towards Improving Skeletons in
  {Eden}}}.
\newblock In \bibinfo{editor}{Gerhard~R. Joubert}, \bibinfo{editor}{Wolfgang~E.
  Nagel}, \bibinfo{editor}{Frans~J. Peters}, \bibinfo{editor}{Oscar~G. Plata},
  \bibinfo{editor}{P.~Tirado} \& \bibinfo{editor}{Emilio~L. Zapata}, editors:
  {\sl \bibinfo{booktitle}{Parallel Computing: Current {\&} Future Issues of
  High-End Computing, Proceedings of the International Conference ParCo 2005,
  13-16 September 2005, Department of Computer Architecture, University of
  Malaga, Spain}}, {\sl \bibinfo{series}{John von Neumann Institute for
  Computing Series}}~\bibinfo{volume}{33}, \bibinfo{publisher}{Central
  Institute for Applied Mathematics, J{\"u}lich, Germany}, pp.
  \bibinfo{pages}{843--850}.

\bibitemdeclare{inproceedings}{wrs/Hidalgo-HerreroVO07}
\bibitem{wrs/Hidalgo-HerreroVO07}
\bibinfo{author}{Mercedes Hidalgo-Herrero}, \bibinfo{author}{Alberto Verdejo}
  \& \bibinfo{author}{Yolanda Ortega-Mall{\'e}n} (\bibinfo{year}{2007}):
  \emph{\bibinfo{title}{Using {Maude} and Its Strategies for Defining a
  Framework for Analyzing {Eden} Semantics}}.
\newblock In \bibinfo{editor}{Sergio Antoy}, editor: {\sl
  \bibinfo{booktitle}{Proceedings of the Sixth International Workshop on
  Reduction Strategies in Rewriting and Programming, WRS 2006, Seattle, WA,
  USA, August 11, 2006}}, {\sl \bibinfo{series}{Electronic Notes in Theoretical
  Computer Science}} \bibinfo{volume}{174(10)}, \bibinfo{publisher}{Elsevier},
  pp. \bibinfo{pages}{119--137}.
\newblock \urlprefix\url{http://dx.doi.org/10.1016/j.entcs.2007.02.051}.

\bibitemdeclare{inproceedings}{entcs/HolenJW09}
\bibitem{entcs/HolenJW09}
\bibinfo{author}{Bjarne Holen}, \bibinfo{author}{Einar~Broch Johnsen} \&
  \bibinfo{author}{Arild Waaler} (\bibinfo{year}{2009}):
  \emph{\bibinfo{title}{Proof Search for the First-Order Connection Calculus in
  {M}aude}}.
\newblock In \bibinfo{editor}{Ro{\c{s}}u}  \cite{wrla/2008}, pp.
  \bibinfo{pages}{173--188}.
\newblock \urlprefix\url{http://dx.doi.org/10.1016/j.entcs.2009.05.019}.

\bibitemdeclare{article}{jcss/Huet81}
\bibitem{jcss/Huet81}
\bibinfo{author}{G{\'e}rard~P. Huet} (\bibinfo{year}{1981}):
  \emph{\bibinfo{title}{A Complete Proof of Correctness of the Knuth-Bendix
  Completion Algorithm}}.
\newblock {\sl \bibinfo{journal}{Journal of Computer and System Sciences}}
  \bibinfo{volume}{23}(\bibinfo{number}{1}), pp. \bibinfo{pages}{11--21}.
\newblock \urlprefix\url{http://dx.doi.org/10.1016/0022-0000(81)90002-7}.

\bibitemdeclare{inproceedings}{rta/KirchnerM95}
\bibitem{rta/KirchnerM95}
\bibinfo{author}{H{\'e}l{\`e}ne Kirchner} \& \bibinfo{author}{Pierre-Etienne
  Moreau} (\bibinfo{year}{1995}): \emph{\bibinfo{title}{Prototyping Completion
  with Constraints Using Computational Systems}}.
\newblock In \bibinfo{editor}{Jieh Hsiang}, editor: {\sl
  \bibinfo{booktitle}{Rewriting Techniques and Applications, 6th International
  Conference, RTA-95, Kaiserslautern, Germany, April 5-7, 1995, Proceedings}},
  {\sl \bibinfo{series}{Lecture Notes in Computer Science}}
  \bibinfo{volume}{914}, \bibinfo{publisher}{Springer}, pp.
  \bibinfo{pages}{438--443}.
\newblock \urlprefix\url{http://dx.doi.org/10.1007/3-540-59200-8_79}.

\bibitemdeclare{inproceedings}{Lescanne89}
\bibitem{Lescanne89}
\bibinfo{author}{Pierre Lescanne} (\bibinfo{year}{1989}):
  \emph{\bibinfo{title}{Completion Procedures as Transition Rules + Control}}.
\newblock In \bibinfo{editor}{J.~D{\'\i}az} \& \bibinfo{editor}{F.~Orejas},
  editors: {\sl \bibinfo{booktitle}{{TAPSOFT'89} Proceedings of the
  International Joint Conference on Theory and Practice of Software
  Development, Barcelona, Spain, March 13-17, 1989}}, {\sl
  \bibinfo{series}{Lecture Notes in Computer Science}} \bibinfo{volume}{351},
  \bibinfo{publisher}{Springer}, pp. \bibinfo{pages}{28--41}.

\bibitemdeclare{article}{lop04}
\bibitem{lop04}
\bibinfo{author}{Rita Loogen}, \bibinfo{author}{Yolanda {Ortega-Mall\'en}} \&
  \bibinfo{author}{Ricardo Pe{\~n}a} (\bibinfo{year}{2005}):
  \emph{\bibinfo{title}{Parallel Functional Programming in {E}den}}.
\newblock {\sl \bibinfo{journal}{Journal of Functional Programming}}
  \bibinfo{volume}{15}(\bibinfo{number}{1}), pp. \bibinfo{pages}{431--475}.

\bibitemdeclare{inproceedings}{entcs/Lucanu09}
\bibitem{entcs/Lucanu09}
\bibinfo{author}{Dorel Lucanu} (\bibinfo{year}{2009}):
  \emph{\bibinfo{title}{Strategy-Based Rewrite Semantics for Membrane Systems
  Preserves Maximal Concurrency of Evolution Rule Actions}}.
\newblock In \bibinfo{editor}{Aart Middeldorp}, editor: {\sl
  \bibinfo{booktitle}{Proceedings of the Eighth International Workshop on
  Reduction Strategies in Rewriting and Programming, WRS 2008, Castle of
  Hagenberg, Austria, July 14, 2008}}, {\sl \bibinfo{series}{Electronic Notes
  in Theoretical Computer Science}} \bibinfo{volume}{237},
  \bibinfo{publisher}{Elsevier}, pp. \bibinfo{pages}{107--125}.
\newblock \urlprefix\url{http://dx.doi.org/10.1016/j.entcs.2009.03.038}.

\bibitemdeclare{incollection}{Marti-OlietMeseguer02b}
\bibitem{Marti-OlietMeseguer02b}
\bibinfo{author}{Narciso Mart{\'\i}-Oliet} \& \bibinfo{author}{Jos{\'e}
  Meseguer} (\bibinfo{year}{2002}): \emph{\bibinfo{title}{Rewriting logic as a
  logical and semantic framework}}.
\newblock In \bibinfo{editor}{Dov~M. Gabbay} \& \bibinfo{editor}{Franz
  Guenthner}, editors: {\sl \bibinfo{booktitle}{Handbook of Philosophical
  Logic, Second Edition, Volume 9}}, \bibinfo{publisher}{Kluwer Academic
  Publishers}, pp. \bibinfo{pages}{1--87}.

\bibitemdeclare{inproceedings}{MartiOlietMeseguerVerdejo04}
\bibitem{MartiOlietMeseguerVerdejo04}
\bibinfo{author}{Narciso Mart{\'\i}-Oliet}, \bibinfo{author}{Jos{\'e} Meseguer}
  \& \bibinfo{author}{Alberto Verdejo} (\bibinfo{year}{2004}):
  \emph{\bibinfo{title}{Towards a Strategy Language for {Maude}}}.
\newblock In \bibinfo{editor}{Narciso Mart{\'\i}-Oliet}, editor: {\sl
  \bibinfo{booktitle}{Proceedings of the Fifth International Workshop on
  Rewriting Logic and its Applications, WRLA 2004, Barcelona, Spain, March
  27-April 4, 2004}}, {\sl \bibinfo{series}{Electronic Notes in Theoretical
  Computer Science}} \bibinfo{volume}{117}, \bibinfo{publisher}{Elsevier}, pp.
  \bibinfo{pages}{417--441}.
\newblock \urlprefix\url{http://dx.doi.org/10.1016/j.entcs.2004.06.020}.

\bibitemdeclare{inproceedings}{entcs/Marti-OlietMV09}
\bibitem{entcs/Marti-OlietMV09}
\bibinfo{author}{Narciso Mart{\'\i}-Oliet}, \bibinfo{author}{Jos{\'e} Meseguer}
  \& \bibinfo{author}{Alberto Verdejo} (\bibinfo{year}{2009}):
  \emph{\bibinfo{title}{A Rewriting Semantics for {M}aude Strategies}}.
\newblock In \bibinfo{editor}{Ro{\c{s}}u}  \cite{wrla/2008}, pp.
  \bibinfo{pages}{227--247}.
\newblock \urlprefix\url{http://dx.doi.org/10.1016/j.entcs.2009.05.022}.

\bibitemdeclare{inproceedings}{iswsa/MerouaniMS10}
\bibitem{iswsa/MerouaniMS10}
\bibinfo{author}{Hamza Merouani}, \bibinfo{author}{Farid Mokhati} \&
  \bibinfo{author}{Hassina Seridi-Bouchelaghem} (\bibinfo{year}{2010}):
  \emph{\bibinfo{title}{Towards formalizing web service composition in
  {Maude}'s strategy language}}.
\newblock In \bibinfo{editor}{Ayman Alnsour} \& \bibinfo{editor}{Shadi
  Aljawarneh}, editors: {\sl \bibinfo{booktitle}{Proceedings of the 1st
  International Conference on Intelligent Semantic Web-Services and
  Applications, ISWSA 2010, Amman, Jordan, June 14-16, 2010}},
  \bibinfo{publisher}{ACM}, p.~\bibinfo{pages}{15}.
\newblock \urlprefix\url{http://doi.acm.org/10.1145/1874590.1874605}.

\bibitemdeclare{article}{tcs/Meseguer92}
\bibitem{tcs/Meseguer92}
\bibinfo{author}{Jos{\'e} Meseguer} (\bibinfo{year}{1992}):
  \emph{\bibinfo{title}{Conditional Rewriting Logic as a Unified Model of
  Concurrency}}.
\newblock {\sl \bibinfo{journal}{Theoretical Computer Science}}
  \bibinfo{volume}{96}(\bibinfo{number}{1}), pp. \bibinfo{pages}{73--155}.
\newblock \urlprefix\url{http://dx.doi.org/10.1016/0304-3975(92)90182-F}.

\bibitemdeclare{inproceedings}{amast/MeseguerB04}
\bibitem{amast/MeseguerB04}
\bibinfo{author}{Jos{\'e} Meseguer} \& \bibinfo{author}{Christiano Braga}
  (\bibinfo{year}{2004}): \emph{\bibinfo{title}{Modular Rewriting Semantics of
  Programming Languages}}.
\newblock In \bibinfo{editor}{Charles Rattray}, \bibinfo{editor}{Savi Maharaj}
  \& \bibinfo{editor}{Carron Shankland}, editors: {\sl
  \bibinfo{booktitle}{Algebraic Methodology and Software Technology, 10th
  International Conference, AMAST 2004, Stirling, Scotland, UK, July 12-16,
  2004, Proceedings}}, {\sl \bibinfo{series}{Lecture Notes in Computer
  Science}} \bibinfo{volume}{3116}, \bibinfo{publisher}{Springer}, pp.
  \bibinfo{pages}{364--378}.
\newblock \urlprefix\url{http://dx.doi.org/10.1007/978-3-540-27815-3_29}.

\bibitemdeclare{article}{tcs/MeseguerR07}
\bibitem{tcs/MeseguerR07}
\bibinfo{author}{Jos{\'e} Meseguer} \& \bibinfo{author}{Grigore Ro{\c{s}}u}
  (\bibinfo{year}{2007}): \emph{\bibinfo{title}{The rewriting logic semantics
  project}}.
\newblock {\sl \bibinfo{journal}{Theoretical Computer Science}}
  \bibinfo{volume}{373}(\bibinfo{number}{3}), pp. \bibinfo{pages}{213--237}.
\newblock \urlprefix\url{http://dx.doi.org/10.1016/j.tcs.2006.12.018}.

\bibitemdeclare{inproceedings}{RosaSeguraVerdejo05}
\bibitem{RosaSeguraVerdejo05}
\bibinfo{author}{Fernando Rosa-Velardo}, \bibinfo{author}{Clara Segura} \&
  \bibinfo{author}{Alberto Verdejo} (\bibinfo{year}{2006}):
  \emph{\bibinfo{title}{Typed Mobile Ambients in {Maude}}}.
\newblock In \bibinfo{editor}{Horatiu Cirstea} \& \bibinfo{editor}{Narciso
  Mart{\'\i}-Oliet}, editors: {\sl \bibinfo{booktitle}{Proceedings of the 6th
  International Workshop on Rule-Based Programming, RULE 2005, Nara, Japan,
  April 23, 2005}}, {\sl \bibinfo{series}{Electronic Notes in Theoretical
  Computer Science}} \bibinfo{volume}{147(1)}, \bibinfo{publisher}{Elsevier},
  pp. \bibinfo{pages}{135--161}.
\newblock \urlprefix\url{http://dx.doi.org/10.1016/j.entcs.2005.06.041}.

\bibitemdeclare{proceedings}{wrla/2008}
\bibitem{wrla/2008}
\bibinfo{editor}{Grigore Ro{\c{s}}u}, editor (\bibinfo{year}{2009}):
  \emph{\bibinfo{title}{Proceedings of the Seventh International Workshop on
  Rewriting Logic and its Applications, WRLA 2008, Budapest, Hungary, March
  29-30, 2008}}. {\sl \bibinfo{series}{Electronic Notes in Theoretical Computer
  Science}} \bibinfo{volume}{238(3)}, \bibinfo{publisher}{Elsevier}.

\bibitemdeclare{inproceedings}{entcs/Santos-GarciaP07}
\bibitem{entcs/Santos-GarciaP07}
\bibinfo{author}{Gustavo Santos-Garc{\'\i}a} \& \bibinfo{author}{Miguel
  Palomino} (\bibinfo{year}{2007}): \emph{\bibinfo{title}{Solving Sudoku
  Puzzles with Rewriting Rules}}.
\newblock In \bibinfo{editor}{Denker} \& \bibinfo{editor}{Talcott}
  \cite{wrla/2006}, pp. \bibinfo{pages}{79--93}.
\newblock \urlprefix\url{http://dx.doi.org/10.1016/j.entcs.2007.06.009}.

\bibitemdeclare{inproceedings}{dcai/Santos-GarciaPV08}
\bibitem{dcai/Santos-GarciaPV08}
\bibinfo{author}{Gustavo Santos-Garc{\'\i}a}, \bibinfo{author}{Miguel Palomino}
  \& \bibinfo{author}{Alberto Verdejo} (\bibinfo{year}{2009}):
  \emph{\bibinfo{title}{Rewriting Logic Using Strategies for Neural Networks:
  An Implementation in {Maude}}}.
\newblock In \bibinfo{editor}{Juan~M. Corchado}, \bibinfo{editor}{Sara
  Rodr{\'i}guez}, \bibinfo{editor}{James Llinas} \&
  \bibinfo{editor}{Jos{\'e}~M. Molina}, editors: {\sl
  \bibinfo{booktitle}{Proceedings of the International Symposium on Distributed
  Computing and Artificial Intelligence, DCAI 2008, University of Salamanca,
  Spain, October 22-24, 2008}}, {\sl \bibinfo{series}{Advances in Soft
  Computing}}~\bibinfo{volume}{50}, \bibinfo{publisher}{Springer}, pp.
  \bibinfo{pages}{424--433}.
\newblock \urlprefix\url{http://dx.doi.org/10.1007/978-3-540-85863-8_50}.

\bibitemdeclare{article}{ulidowski}
\bibitem{ulidowski}
\bibinfo{author}{Irek Ulidowski} \& \bibinfo{author}{Iain Phillips}
  (\bibinfo{year}{2002}): \emph{\bibinfo{title}{Ordered {SOS} Process Languages
  for Branching and Eager Bisimulations}}.
\newblock {\sl \bibinfo{journal}{Information and Computation}}
  \bibinfo{volume}{178}, pp. \bibinfo{pages}{180--213}.

\bibitemdeclare{inproceedings}{Visser01}
\bibitem{Visser01}
\bibinfo{author}{Eelco Visser} (\bibinfo{year}{2001}):
  \emph{\bibinfo{title}{Stratego: A Language for Program Transformation Based
  on Rewriting Strategies}}.
\newblock In \bibinfo{editor}{Aart Middeldorp}, editor: {\sl
  \bibinfo{booktitle}{Rewriting Techniques and Applications, 12th International
  Conference, RTA 2001, Utrecht, The Netherlands, May 22-24, 2001,
  Proceedings}}, {\sl \bibinfo{series}{Lecture Notes in Computer Science}}
  \bibinfo{volume}{2051}, \bibinfo{publisher}{Springer}, pp.
  \bibinfo{pages}{357--362}.

\bibitemdeclare{inproceedings}{Visser04}
\bibitem{Visser04}
\bibinfo{author}{Eelco Visser} (\bibinfo{year}{2004}):
  \emph{\bibinfo{title}{Program Transformation with {Stratego/XT}: Rules,
  Strategies, Tools, and Systems in {StrategoXT-0.9}}}.
\newblock In \bibinfo{editor}{C.~Lengauer}, editor: {\sl
  \bibinfo{booktitle}{Domain-Specific Program Generation}}, {\sl
  \bibinfo{series}{Lecture Notes in Computer Science}} \bibinfo{volume}{3016},
  \bibinfo{publisher}{Springer}, pp. \bibinfo{pages}{216--238}.

\end{thebibliography}

\end{document}